\RequirePackage[2020-02-02]{latexrelease}
\documentclass[aip,reprint]{revtex4-1}
\usepackage[pdftex]{graphicx} 
\usepackage{float}
\usepackage{amsmath,color}
\usepackage{subcaption}
\usepackage{dcolumn}
\usepackage{soul}
\draft 

\newcommand{\dd}{\mathrm{d}}

\begin{document}


\title{Super-Fermi Acceleration in Multiscale MHD Reconnection}

\author{Stephen Majeski}
 \email{smajeski@princeton.edu}
\author{Hantao Ji}%
 \altaffiliation{Also at Princeton Plasma Physics Laboratory}
\affiliation{%
 Department of Astrophysical Sciences,
 Peyton Hall, 4 Ivy Lane,
Princeton University, Princeton, NJ 08544
}%

\date{\today}

\begin{abstract}

We investigate the Fermi acceleration of charged particles in 2D MHD anti-parallel plasmoid reconnection, finding a drastic enhancement in energization rate $\dot{\varepsilon}$ over a standard Fermi model of $\dot{\varepsilon} \sim \varepsilon$. The shrinking particle orbit width around a magnetic island due to $\vec{E}\times\vec{B}$ drift produces a $\dot{\varepsilon}_\parallel \sim \varepsilon_\parallel^{1+1/2\chi}$ power law with $\chi \sim 0.75$. The increase in the maximum possible energy gain of a particle within a plasmoid due to the enhanced efficiency increases with the plasmoid size, and is by multiple factors of 10 in the case of solar flares and much more for larger plasmas. Including effects of the non-constant $\vec{E}\times\vec{B}$ drift rates leads to further variation of power law indices from $\gtrsim 2$ to $\lesssim 1$, decreasing with plasmoid size at the time of injection. The implications for energetic particle spectra are discussed alongside applications to 3D plasmoid reconnection and the effects of a guide field.
\end{abstract}


\maketitle 



\section{Introduction}

Energy conversion in magnetic reconnection is pivotal to understanding reconnection's role throughout the Universe~\cite{yamada10,ji11,ji22}. In solar flares, estimates have found as much as half of electrons being energized to non-thermal energies~\cite{Emslie2005,krucker10}. Moreover, within the solar wind and the earth's magnetotail, electron acceleration and power law energy spectra are often found associated with plasmoids and compressing or merging flux ropes~\cite{oieroset02,chen09,khabarova2016,zhao2018,zhao2019}. Recent years have seen considerable effort to explain these observations, focusing on three leading mechanisms during reconnection: direct acceleration by reconnection electric field~\cite{zenitani01,uzdensky11,chien22} or by localized instances of magnetic field-aligned electric fields~\cite{egedal_daughton_le_2012}, betatron acceleration due to field compression while conserving particle magnetic moments~\cite{Hoshino2001,borovikov_2017,hakobyan_2021}, and Fermi acceleration by ``kicks" from the motional electric field within islands~\cite{drake06,dahlin14,guo_14,Montag2017}. Fermi acceleration operates primarily in multiscale, or plasmoid, reconnection which is thought to be pervasive from solar flares to magnetospheric substorms to accretion disks~\cite{zweibel_yamada_2009,sironi2014,ripperda_17,philippov2019}. In these environments, it takes place within the large volume of magnetic islands which pervade plasmoid-unstable current sheets~\cite{guo15}. A unique characteristic of Fermi acceleration which makes it particularly promising for explaining power law distributions, is that the acceleration rate is itself a power law in energy~\cite{drake06}. This has led to simulations finding Fermi-generated power law distributions over a range of Lundquist numbers, Lorentz factors, guide fields, and more~\cite{guo_2019, ball19}.


Analytical estimates of first-order Fermi acceleration are frequently based off of the seminal work of Drake et al, which found that the particle acceleration rate is linear in the particle energy, $\dot{\varepsilon} \sim \varepsilon$ (in what follows we will refer to acceleration rate power law indices with $p$, \textit{i.e.} $\dot{\varepsilon} \sim \varepsilon^p$)~\cite{drake06}. Note that we are concerned here in this work only with first-order Fermi acceleration which should not be confused with less efficient, second-order, or stochastic, Fermi acceleration. Other approaches have described Fermi acceleration in more MHD-like plasmoid mergers via conservation of the bounce invariant $J_\parallel$~\cite{Montag2017,hakobyan_2021}. Building off of these concepts, energetic particle spectral indices over a range of values larger than $1$ have been explained through a combination of Fermi acceleration, various drifts, and particle-loss processes~\cite{guo_14,guo15}. Efforts have also been made to implement the kinetic physics of Fermi acceleration without resolving small scales~\cite{arnold2019}. Unfortunately, most \textit{analytical} particle acceleration studies are developed to explain the results of kinetic simulations which are computationally limited in the scale separation between large MHD magnetic islands and the Larmor radius ($\rho_L$) of accelerating particles. Yet many astrophysical systems showing promise as a source for energetic particles are deep within the MHD regime~\cite{ji11,guo20}. Such lack of scale separation leads to difficulty in capturing effects like the conservation of adiabatic invariants, increasing loss rates from magnetic islands through pitch-angle scattering~\cite{lgsk_19, dds_15}. Additionally, for lower energy but still weakly collisional particles, their bounce motion may not be fast enough to assume conservation of $J_\parallel$. We therefore propose a new model of Fermi-like acceleration in 2D MHD anti-parallel reconnection, which focuses on systems with large scale separation between thermal particle Larmor radii and plasmoid sizes. With the aid of guiding-center test particle simulations, we find that enhanced particle confinement to compressing magnetic field lines yields an $\mathcal{O}(1)$ correction to the linear Fermi power law index $p=1$.

\subsection{Linear Fermi acceleration}

\begin{figure}[h]
\centering
\includegraphics[width=0.48
\textwidth]{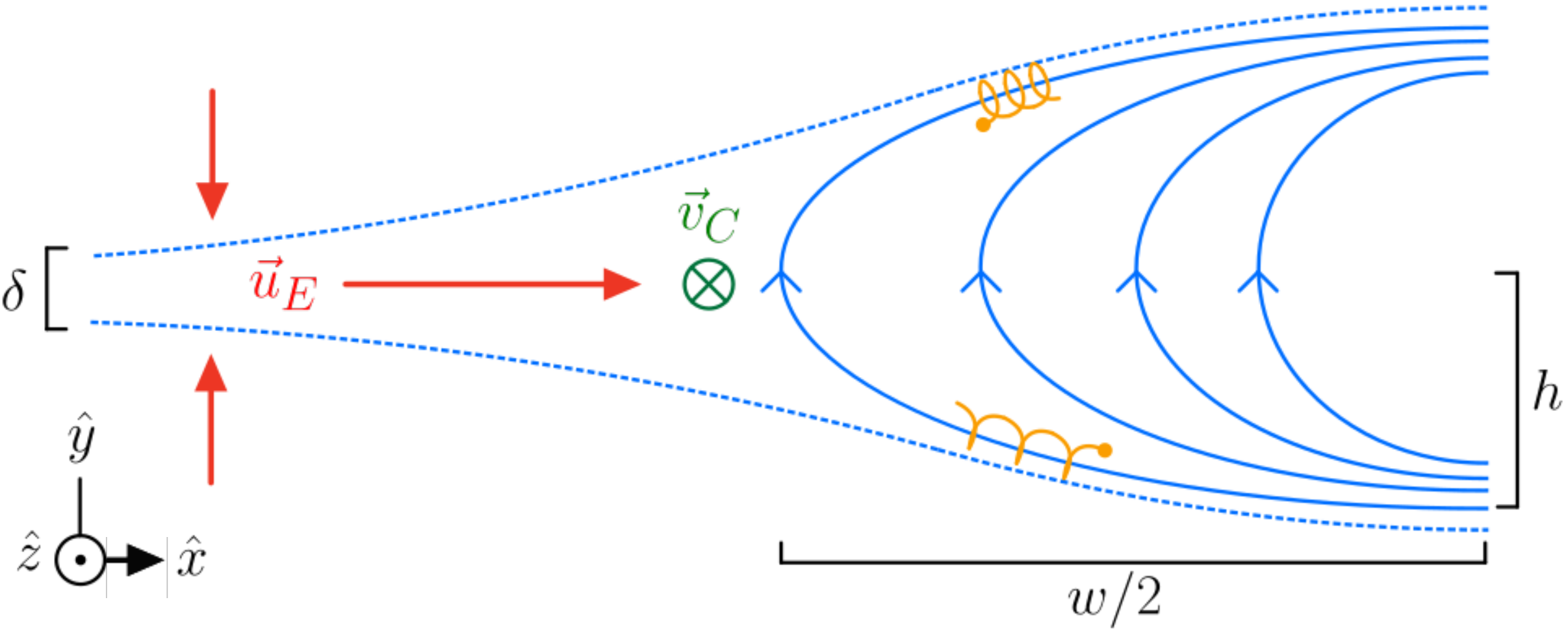}
\caption{Diagram of Fermi acceleration process. Blue lines represent the magnetic field (separatrix dashed), and the curvature drift is given for a positively charged particle.}\label{fig:diagram}
\end{figure}

Consider a plasmoid embedded in a current sheet undergoing 2D anti-parallel MHD reconnection with electric and magnetic fields $\vec{E}$ and $\vec{B}$, respectively. Away from the x-point, the dominant electric field component is the motional field which drives the ``E cross B" drift $\vec{u}_E = c\vec{E} \times \vec {B}/B^2$, which, along with all other electric field components, is out-of-plane in this setup~\cite{parker_57}. If a magnetized particle within a plasmoid is to gain energy, it must experience net motion along this electric field, in this case via guiding center drift. The only drift in this circumstance satisfying this constraint is the curvature drift $\vec{v}_C$. Note that we've assumed drifts arising from explicit time dependence can be neglected, unlike those resulting from particle motion along gradients in $\hat{b}$. This is due to the slow nature of the MHD background compared to the relatively fast motional time derivatives experienced by high energy (and importantly super-Alfv\'enic) particles. Figure~\ref{fig:diagram} shows the process of Fermi acceleration in such a setup. 

As a magnetized, $\mu=mv_{\perp}^2/2B$ conserving particle travels along a field line within the plasmoid (with $m$ the particle mass and $v_\perp$ the particle velocity perpendicular to $\vec{B}$), it enters a narrow region (with respect to the orbit's vertical height $h$) near the neighboring x-point of thickness $\Delta$ which is defined by a large value of the curvature of the magnetic field. This region is generally somewhat larger than the current sheet thickness $\delta$, but approaches that value with increasing proximity to the x-point. The magnetic tension in this high curvature region drives the magnetic field to rapidly straighten out, therefore within $\Delta$ the $\vec{E}\times\vec{B}$ drift velocity is also large. In 2D anti-parallel reconnection, the $\vec{E}\times\vec{B}$ associated electric field and the curvature drift are aligned, therefore the parallel energy of the particle is increased according to $\dot{\varepsilon}_\parallel = 2q\vec{E} \cdot \vec{v}_C/m$, where $\varepsilon_\parallel \doteq v_\parallel^2$, and 
%
%
\begin{equation}
    \vec{v}_C = \frac{m \varepsilon_\parallel}{q B} \hat{b} \times (\hat{b}\cdot \nabla \hat{b}) \approx - \frac{2m\varepsilon_\parallel}{\Delta q B} \hat{z}. \label{eq:curve}
\end{equation}
Note we have assumed here that the gradient scale of $\hat{b}$ is approximately $\Delta/2$. The increase in $\varepsilon_\parallel$ gained by the particle during its transit of $\Delta$ is then estimated as $\dot{\varepsilon}_\parallel \Delta/\sqrt{\varepsilon_\parallel} \approx  4 \langle u_E \rangle_\Delta \sqrt{\varepsilon_\parallel}$, with $\langle \rangle_\Delta$ representing the average over the narrow layer $\Delta$. We have also used $|u_E|=|E/B|$, and assumed that $v_\parallel \gg |u_E|$ in keeping with Drake et al~\cite{drake06}. This process occurs each time the particle transits the island width $w$, which takes a time $d t_w \approx w/\sqrt{\varepsilon_\parallel}$, yielding the linear Fermi acceleration rate:
\begin{equation}
    \biggl(\frac{\dd \varepsilon_\parallel}{\dd t}\biggr)_F \approx 4 \langle u_E \rangle_\Delta \frac{\varepsilon_\parallel}{w}. \label{eq:fermi1}
\end{equation}
This expression is identical in appearance to that of Drake et al, with key differences in meaning~\cite{drake06}. The assumptions under which this equation was derived are MHD without a guide field, not kinetic, meaning no $E_\parallel$ or Hall magnetic field component is present. Equation \eqref{eq:fermi1} has the appearance of being linear in energy, however we will show that during a particle's acceleration $\langle u_E \rangle_\Delta$ and $w$ are not constant, leading to deviation from the linear dependence.

\section{Test Particle Simulations}

\begin{figure}[h]
\vspace{2ex}
\hspace*{-3ex}   
\includegraphics[width=0.47\textwidth]{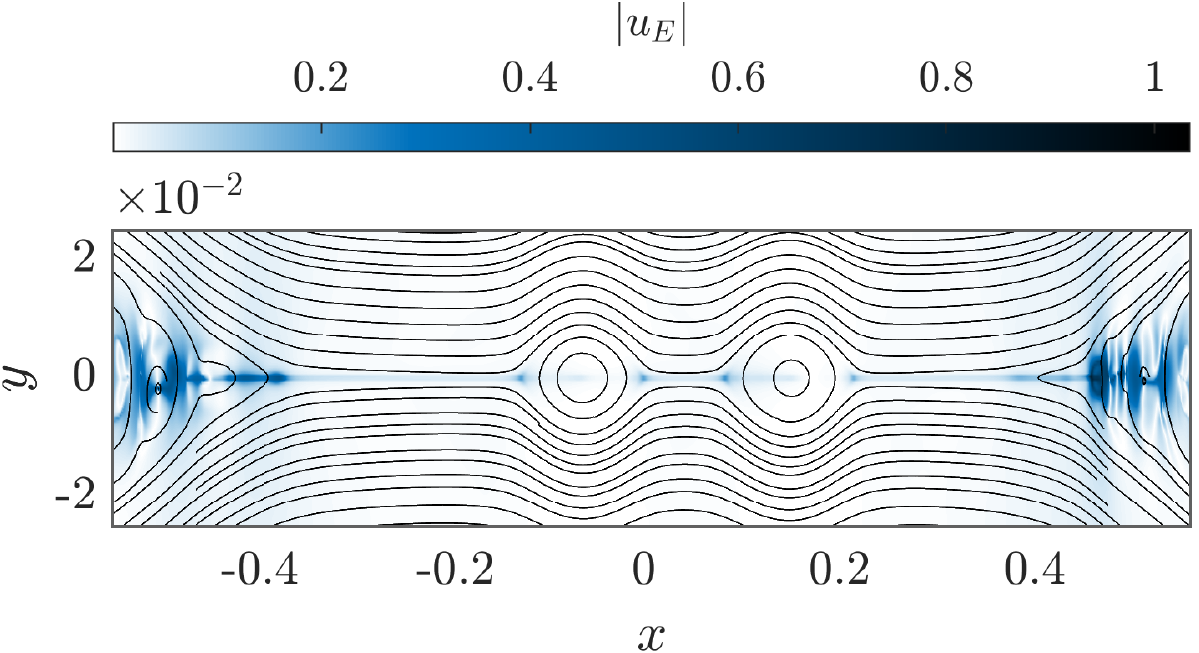}
\caption{Plot of full reconnecting current sheet in the simulation used, at $t=2.9L/v_A$. $|u_E|=E/B$ is shown with streamlines of the magnetic field overlaid. The reconnection layer is formed by vertically merging two large initial magnetic islands.~\cite{huang_bhattacharjee_2010}}\label{fig:pmoids}
\end{figure}

To investigate possible variation of $\langle u_E\rangle_\Delta$ and $w$ in Eq.\eqref{eq:fermi1}, we performed guiding center simulations of test particles in a plasmoid reconnection scenario. To be precise, we solved the following set of simplified non-relativistic guiding-center equations
~\cite{northrop_61, ripperda_17}
\begin{subequations}
\begin{gather}
   \frac{\dd \vec{R}}{\dd t} = v_\parallel \hat{b} + \vec{u}_E \label{eq:R}\\
   \frac{\dd v_\parallel}{\dd t} =  \frac{q}{m} E_\parallel + \vec{u}_E \cdot \left[\left(v_\parallel \hat{b} + \vec{u}_E\right) \cdot \nabla \hat{b}\right] - \frac{\mu}{m}\hat{b} \cdot \nabla B \label{eq:accel},
\end{gather}\label{eq:gce}
\end{subequations}
by an adaptive time step 2nd order-accurate midpoint method~\cite{NumRec}, using time-evolving background data from a 2D MHD simulation~\cite{huang_bhattacharjee_2012, huang_bhattacharjee_2010}. The code which provided the background fields solves the fully-compressible resistive MHD equations via finite differences with a five point spatial stencil and second-order trapezoidal leapfrog time stepping~\cite{guzdar_1993}. These guiding center equations have been simplified assuming that the time dependent drifts are weak due to the slow nature of the MHD background compared to the motional time dependence of super-Alfv\'enic particles. Out-of-plane motion of the guiding center (but not out-of-plane acceleration) is ignored given the 2D symmetry, and in the MHD simulation data used, $E_\parallel=0$. An example snapshot of $u_E$ from the simulation is shown in figure \ref{fig:pmoids}. Note that when interpreting the magnitude of $u_E$, the density and magnetic field away from the current sheet in this simulation approach $\rho_0 = B_0 = 1$ in dimensionless numerical units. The spatial grid size is 2000 ($x$) by 4000 ($y$) and time outputs are available at intervals of one-tenth of the primary current sheet Alfv\'en time (for context, the snapshot fig. \ref{fig:pmoids} shows a zoomed-in portion of the grid which is 1000 by 100 cells). As a result, linear interpolation from the MHD grid to the particle's time and position is used. The background plasma beta is $\beta=1$, with a uniform Lundquist number of $S = 10^5$.
The adaptive particle time step is calculated as a fraction (CFL number) of the simulation grid cell-crossing time for the particle's velocity, including the $\vec{E}\times\vec{B}$ drift. In all calculations shown the CFL number is set to 0.1. In the MHD simulation, two plasmoids form, which eventually begin to merge at $t=3.6L/v_A$, where $L$ is the $x$-extent of the simulation domain~\cite{huang_bhattacharjee_2010}. As a result, we limit our study to pre-merger times in the simulation to avoid the further complication of acceleration at the secondary current sheet. The initial particle velocity is set to $v_\parallel = 20v_A$ for the purpose of ensuring the small $\Delta \varepsilon_\parallel$ approximations holds, however no significant difference was noticed in runs where the initial velocity was $10v_A$ or $5v_A$. How the particles are initially energized relates to the problem of injection, which is a very active area of study but beyond the scope of this work~\cite{ball19,Killian2020,French22,sironi22}. The perpendicular velocity of particles was set to $v_\perp = v_A$ and generally plays little role unless $v_\perp \sim v_\parallel$, which leads to particle trapping at the island edge.
\begin{figure}[h]
\centering
\includegraphics[width=0.45\textwidth]{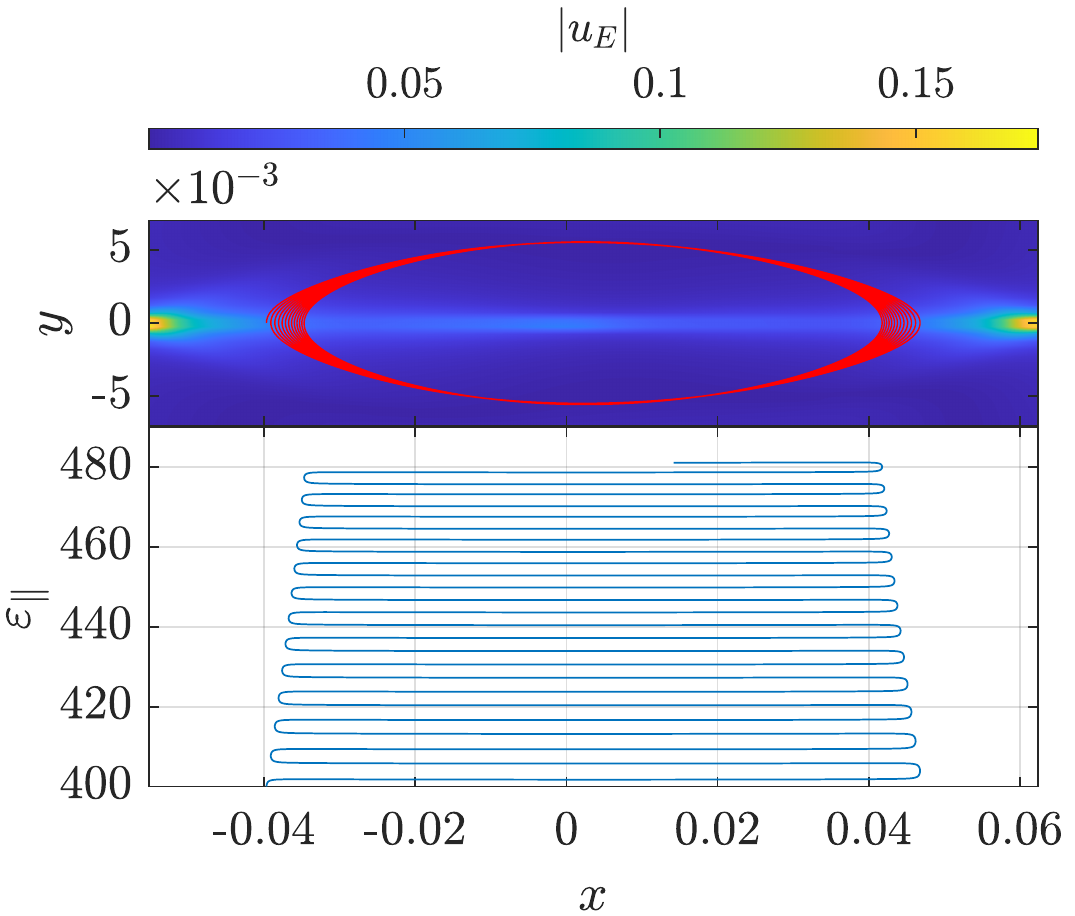}
\caption{Particle orbit path for injection at $t=2.7L/v_A$, overlaid on initial magnitude of $\vec{u}_E$. Total time of integration is $t=0.1L/v_A$.}\label{fig:orbit}
\end{figure}

An example test particle orbit is shown in Fig.\ref{fig:orbit} with the initial $u_E$ field that it experienced, for a total evolution time of $t=0.1L/v_A$. $\vec{u}_E$ is seen to be limited to a narrow central section which is approximately uniform in width, and peaks in magnitude at the reconnection outflow. In the following, we will refer to the effective plasmoid width $w_p$ as the distance between the two maxima of $u_E$. The particle orbit shows a steady decrease in width $w$, also visible in the plot of $\varepsilon_\parallel$ versus $x$-position. Conversely, there is no comparable change in $h$ (given in Fig.\ref{fig:diagram}). This particle was injected with an initial orbit width $w_0 = 3w_p/4$ at $t=2.7L/v_A$, roughly $0.5L/v_A$ after the plasmoid became nonlinear (which we consider here as the point when the plasmoid's vertical extent $h_p$ exceeds the current sheet thickness, \textit{i.e.} $h_p \gtrsim \delta$). From this data, we fit a power law to the bounce average of $\dot{\varepsilon}_\parallel/\langle u_E\rangle_\Delta$, finding an exponent of 1.77, rather than the predicted value of 1 from linear Fermi acceleration. Alternatively, fitting a power law to $\dot{\varepsilon}_\parallel w/\langle u_E\rangle_\Delta$ yields a power law index of 1.08. This suggests that the non-constancy of $w$ may account for the disagreement with the linear Fermi prediction. In section \ref{sec:orbwidth}, we will therefore attempt to describe the nature of the $\varepsilon_\parallel$-$w$ relationship to more generally predict the energy dependence of $\dot{\varepsilon}_\parallel/\langle u_E\rangle_\Delta$.

\subsection{Orbit width correction}\label{sec:orbwidth}

\begin{figure}[h]
\centering
\includegraphics[width=0.46\textwidth]{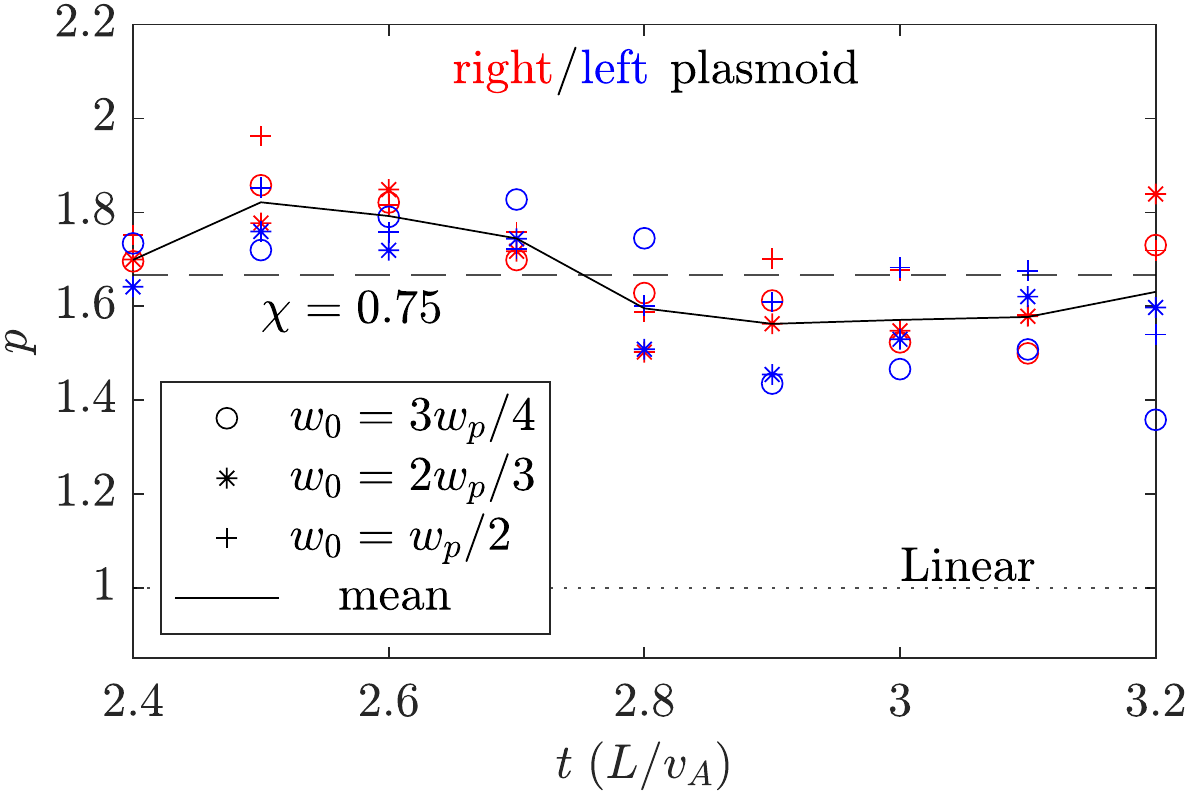}
\caption{Power law fits to test particle data for $0.3L/v_A$ of evolution time, with the super-Fermi exponent assuming $\chi = 0.75$.}\label{fig:scatter}
\end{figure}

The electric field which does work on curvature-drifting particles in our linear Fermi acceleration calculation results from the field line motion that compresses plasmoids. Naturally then, as particles gain energy from the Fermi acceleration process, the closed field lines they are bound to shrink in width (also see Fig. \ref{fig:orbit}):
\begin{equation}
    \frac{\dd w}{\dd t} = -2\langle u_E\rangle _{pk}.\label{eq:drift}
\end{equation}
Here $\langle u_E \rangle_{pk}$ is the \textit{peak} value (not $\Delta$-averaged) of $u_E$ experienced by the particle as it transits \textit{both} sides of the island (averaged between the left and right). This distinction is due to the fact that generally $\langle u_E \rangle_{pk}$ occurs in the locations of highest curvature, \textit{i.e.} at the extreme edges of the island. Therefore the rate at which these extremes contract sets the rate of change of $w$. This peak value is generally slightly larger than $\langle u_E \rangle_\Delta$, and we will assume that the ratio $\langle u_E \rangle_\Delta/ \langle u_E \rangle_{pk} = \chi$ is approximately constant over the period during which a particle is accelerated. Qualitatively, this is an assumption that as long as the outflow $u_E$ remains somewhat laminar, it will maintain a similar functional form as it expands into the plasmoid (this will be checked via the constancy of $\chi$ within figure \ref{fig:scatter}). We then substitute for $\langle u_E\rangle _\Delta$ in Eq.\eqref{eq:fermi1} allowing for the determination of $w(\varepsilon_\parallel)$:
\begin{equation}
    \frac{\dd w}{\dd \varepsilon_\parallel} = -\frac{1}{2\chi}\frac{w}{\varepsilon_{\parallel}}, \qquad w = w_0 \left(\frac{\varepsilon_{\parallel}}{\varepsilon_{\parallel 0}}\right)^{-1/2\chi},\label{eq:evw}
\end{equation}
leading to an enhanced power law acceleration rate:
\begin{equation}
    \biggl(\frac{\dd \varepsilon_\parallel}{\dd t}\biggr)_{SF} \approx 4\chi \frac{\varepsilon_{\parallel 0}}{w_0} \langle u_E \rangle_{pk} \left(\frac{\varepsilon_{\parallel}}{\varepsilon_{\parallel 0}}\right)^{1+1/2\chi}.\label{eq:super}
\end{equation}
The subscript ``SF" has been added for ``Super-Fermi", because the orbit-width correction exclusively leads to stronger energization over the linear expression. Additionally, although $\chi$ is assumed to be constant it can vary somewhat due to minute details of the plasmoid structure. We will therefore make use of simulation data to provide a reasonable estimate. For ultra-relativistic particles which have $v_{\parallel} \approx c$ (or $\gamma \gg 1$), the super-Fermi acceleration rate is
\begin{equation}
    \biggl(\frac{\dd \varepsilon_\parallel}{\dd t}\biggr)_{SF,UR} \approx 2\chi \frac{\varepsilon_{\parallel 0}}{w_0} \langle u_E \rangle_{pk} \left(\frac{\varepsilon_{\parallel}}{\varepsilon_{\parallel 0}}\right)^{1+1/\chi}.\label{eq:superrel}
\end{equation}
where the change $\varepsilon_\parallel \approx \gamma m_0 c^2$ is made but all other variables carry the same meaning~\cite{ripperda_17}. The missing factor of 2 is a result of the ultra-relativistic particle velocity remaining approximately constant. In the non-relativistic case, $\dd_t\varepsilon_\parallel = 2v_\parallel \dd_tv_\parallel$, while ultra-relativistically $\dd_t \varepsilon_\parallel = \dd_t (pc) \approx v_\parallel \dd_t p$, with no additional factor of two. Equation \eqref{eq:superrel} indicates that the ultra-relativistic orbit-width power law correction is twice that of the non-relativistic version.

 To complete equation \ref{eq:super}, a suitable estimate for $\chi$ is needed. Being the result of a plasmoid's internal structure, its precise value will be unique to each plasmoid, although many plasmoids within a given current sheet may possess similar internal structures given their shared origin. To obtain this estimate, a single test particle was evolved inside a plasmoid for a time of $L/v_A$, and $\langle u_E \rangle_\Delta/\langle u_E \rangle_{pk}$ was calculated for 627 transits of the acceleration regions. An acceleration region is detected numerically as the time frame during which a particle experiences a $\Delta \varepsilon_\parallel$ per time step of at least 25\% of the maximum value during the same kick. This yielded a mean $\chi$ of 0.75 ($p=1.67$ non-relativistically, $p=2.33$ ultra-relativistically), which will serve as our fiducial value henceforth. This inferred power law of $p=1.67$ agrees with the value of 1.77 from the particle in figure \ref{fig:orbit} to within 6\%.

To test Eq.~\eqref{eq:super} more broadly we performed a survey of simulations to calculate the particle acceleration rate power law with varying injection times and locations, shown in Fig.\ref{fig:scatter}. 
Test particles were injected into both plasmoids, at 3 different initial orbit widths ($3w_p/4$, $2w_p/3$, $w_p/2$), and 9 different time steps in the MHD simulation (each separated by $0.1L/v_A$). A total of 53 orbit-width corrected power law indices were calculated after $0.3L/v_A$ of evolution time for each particle, excluding one particle which reached the center of its respective plasmoid before the end of the simulation. This interval was chosen to maximize the evolution time that a power law could be fit to, while also providing sufficient data points to determine whether $p$ varies significantly with time (as it is limited by the eventual merger of the right and left plasmoids). To remove the effect of the varying $u_E$, power laws are fit to $\dot{\varepsilon}_\parallel/\langle u_E \rangle_{pk}(t,w)$ to determine the index $p$, rather than just $\dot{\varepsilon}_\parallel$. Both plasmoids are similar in size at each time step, therefore their power law indices are counted together, yet they can be distinguished by the color of their data points' markers. The fiducial power law $p=1.67$ predicted $\chi=0.75$ is shown as a black dashed line, while the linear Fermi prediction is shown as a black dotted line. The average measured power law agrees with the fiducial index to within 9\% at all times, with a time-averaged $p=1.66$. They also demonstrate importantly that there is no net trend in the orbit-width corrected index $p$ with the size of the plasmoid, suggesting that our assumption of constant $\chi$ is suitable. We will however show that the implied power law does not remain constant when including the variation in $\langle u_E \rangle_{pk}$. Regardless of our choice of fiducial $\chi$, the expected lower limit on possible power law indices is 1.5, which is obeyed reasonably well, with a maximum $p$ of 2 suggesting that $\chi$ is generally at least 0.5. The fluctuations seen in our measured power law indices may be the result of weakly time dependent $\chi$, and/or variations in calculated $\langle u_E \rangle_{pk}$ when removing its dependence from $\dot{\varepsilon}_{\parallel}$ numerically.

\subsection{Space- and time-varying $\pmb{\vec{E}\times\vec{B}}$}

The effects of variation in $\langle u_E\rangle_{pk}$ are, unlike the orbit-width correction, highly dependent on the time of injection through the evolution of the plasmoid structure. Such effects have been studied in turbulence, highlighting the relationship between $\vec{u}_E$ gradients and stochastic Fermi acceleration~\cite{lemoine22}. Similarly, previous reconnection-focused work has addressed this issue in the more circular pressure-balanced cores of large plasmoids~\cite{hakobyan_2021}, however we are concerned with the highly elongated outer region of a plasmoid. Without knowledge of the internal plasmoid structure, we have no analytical means by which to determine the modification to the acceleration rate power law by $u_E$. However, a trend in the behavior is  identifiable through a survey of particle injection times when plasmoids possess a variety of sizes/fluxes. Here, we will investigate these effects specifically within the left plasmoid.
\begin{figure}[h]
\centering
\includegraphics[width=0.46
\textwidth]{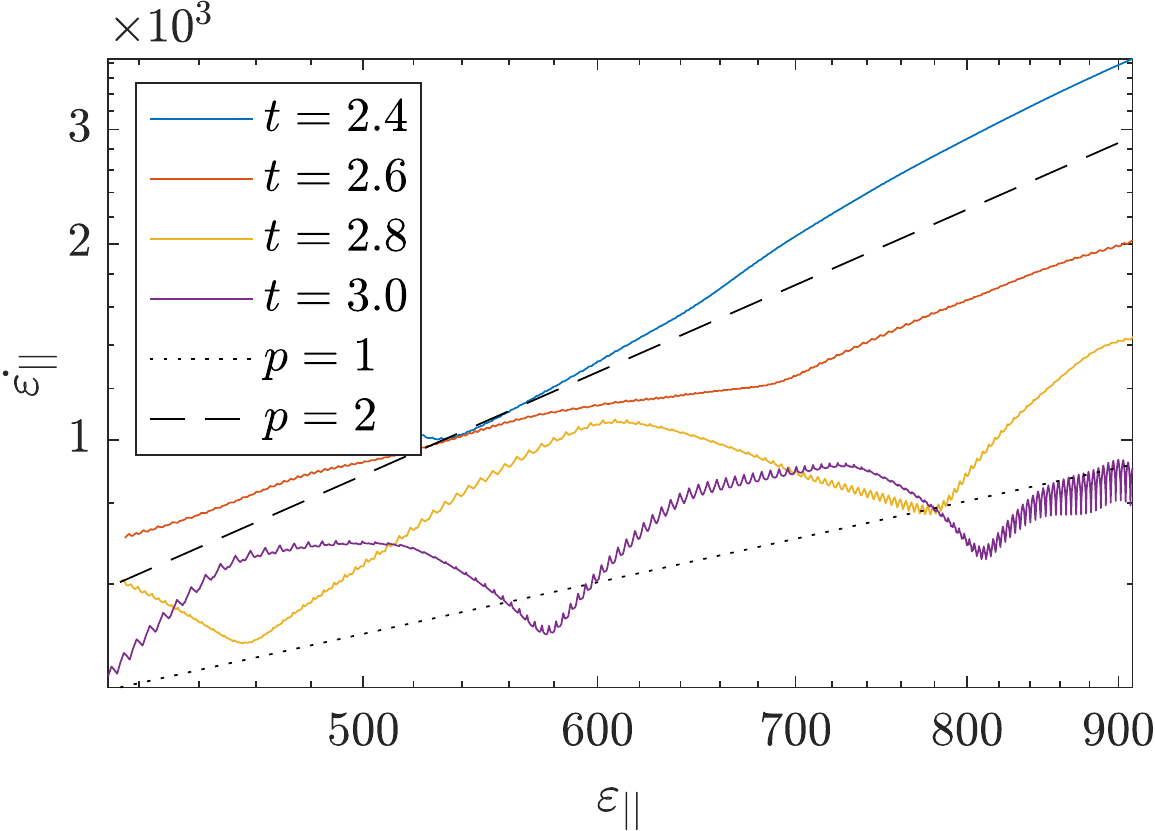}
\caption{Acceleration rates of test particles injected at various times throughout a plasmoid's life. Initial orbit width is $w_0 = 2w_p/3$ for each time.}\label{fig:sizedep}
\end{figure}
In Fig.\ref{fig:sizedep}, particles were injected with $w_0 = 2w_p/3$, and all were evolved for at least $0.5L/v_A$. The trend in $\dot{\varepsilon}_\parallel$ demonstrates that as particles are injected later and later into a plasmoid, the effective power law index of their acceleration decreases. For nearly linear plasmoids the power law index can be larger than 2, while for large nonlinear plasmoids the power law index is able to drop below the linear Fermi rate. This variation occurs due to both the spatial and temporal dependence of $\langle u_E \rangle_{pk}$. The evolution of an $x$-slice of $|u_E|$ within the reconnection layer for the left plasmoid is shown in Fig.\ref{fig:pmoid}.~\cite{huang_bhattacharjee_2010}
\begin{figure}[h]
\centering
\includegraphics[width=0.48
\textwidth]{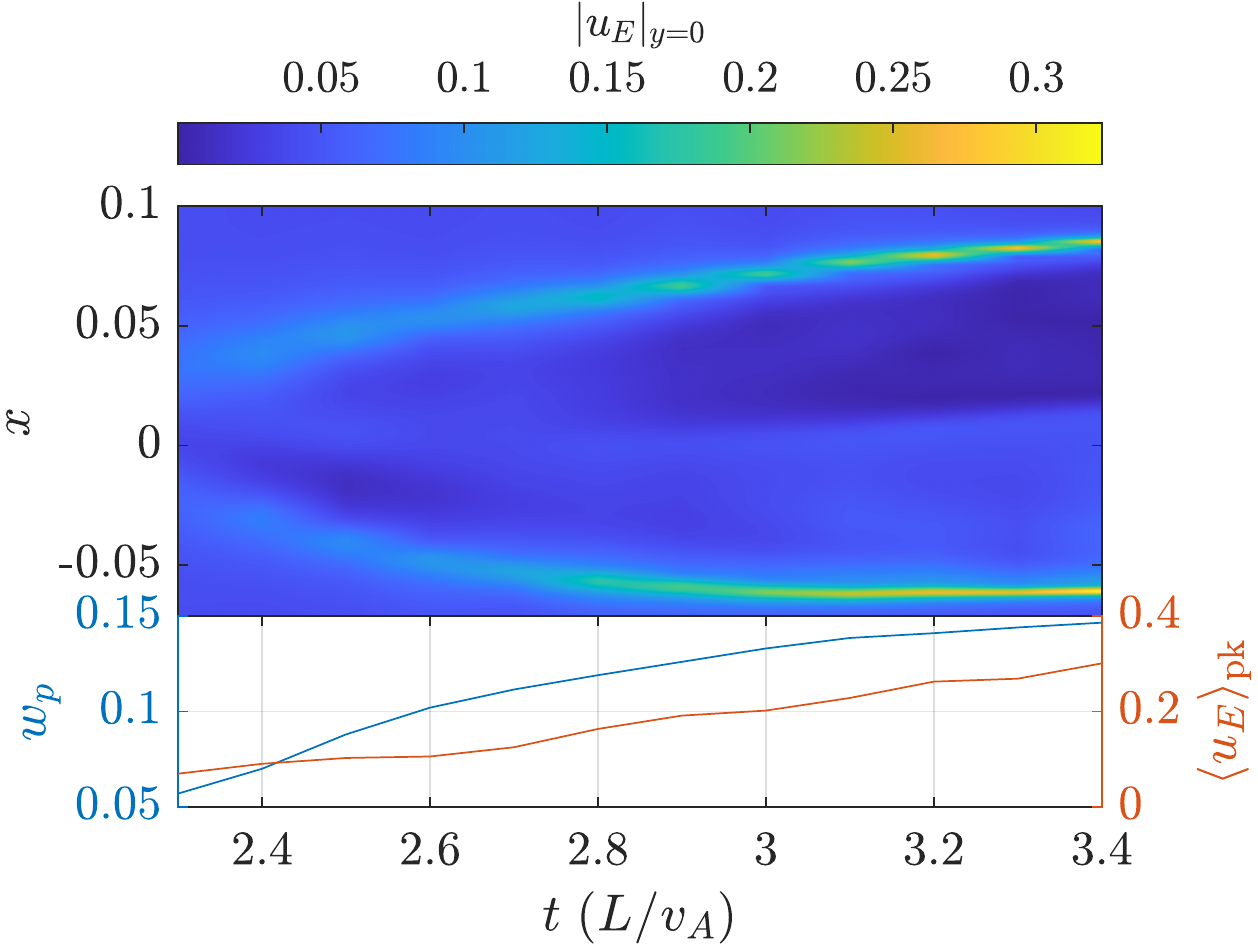}
\caption{Diagram of the evolution of the left plasmoid's $|u_E|$ at $y$=1e-3, with $\langle u_E \rangle_{pk}$ and $w_p$ versus time highlighted below.}\label{fig:pmoid}
\end{figure}
Strong negative gradients are visible in the magnitude of $u_E$ as one moves inward from the edges of the plasmoid. These spatial gradients relate to $\varepsilon_\parallel$ through Eq.\eqref{eq:evw}, and as a particle drifts inwards the field $u_E$ that it experiences decreases, reducing the effective power law of the acceleration. These gradients become more pronounced as the plasmoid grows, further reducing the power law index of acceleration. The modeling of these gradients is a complex problem of nonlinear plasmoid structure, however, qualitatively they may be expected to appear through the conservation of particle flux, or via the buildup of magnetic flux within the plasmoid. While the reconnection outflow expands into the plasmoid from the x-point, the cross sectional area which the flow penetrates increases, causing the flow velocity to decrease. As a plasmoid grows, the area the outflow expands into becomes increasingly large, and therefore the inward gradient becomes more severe. In terms of flux buildup, as a plasmoid grows the magnetic pressure within the island increases and larger values of the magnetic field's strength push closer to the x-points. Given that $u_E \sim 1/B$, this creates negative gradients in $u_E$ which grow in time as the magnetic flux builds up within the island. Concurrently, the peak value of $|u_E|$ grows with the size of the plasmoid. However, this is strictly limited to the neighborhood of the outflow.

\section{Discussion}

\begin{figure}[h]
\centering
\includegraphics[width=0.46\textwidth]{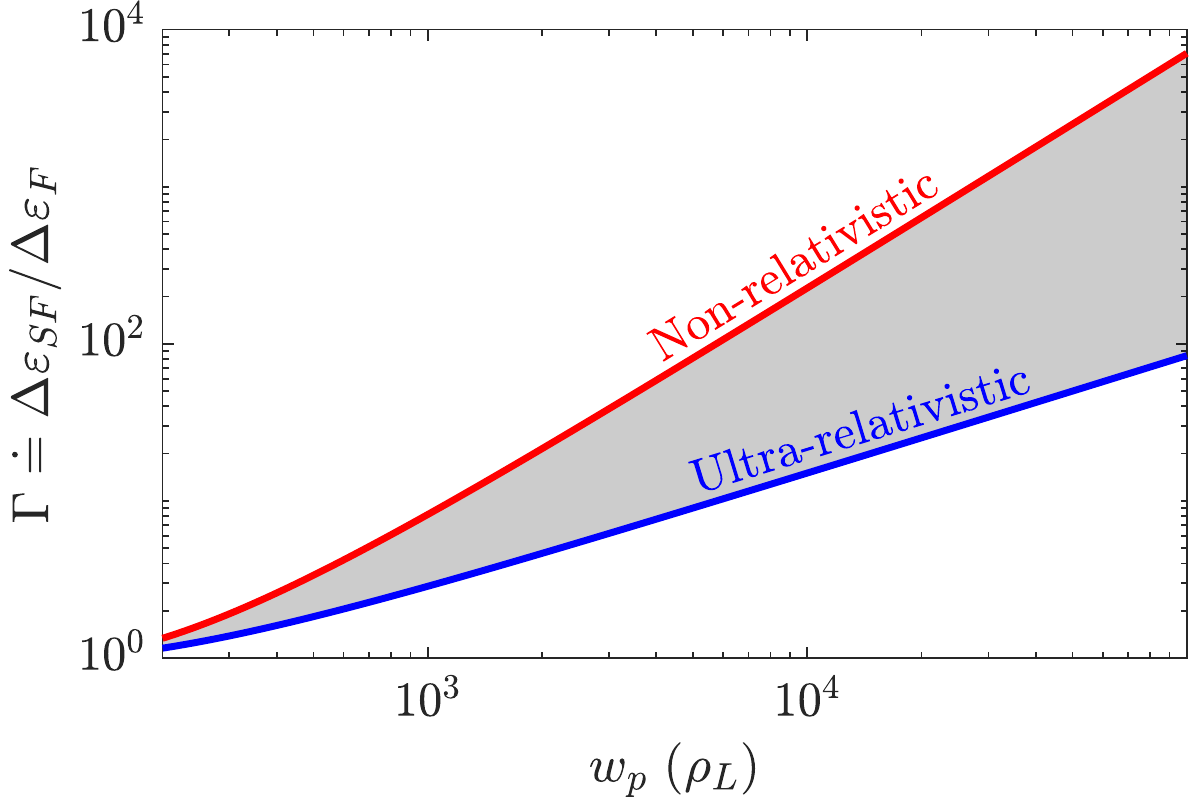}
\caption{Ratio of Super-Fermi to linear Fermi maximum possible total energy gain for a single plasmoid given $w_p$, assuming a particle can only drift inward until roughly $w = 100 \rho_L$.}\label{fig:compare}
\end{figure}

 To quantify the difference between linear and Super-Fermi acceleration, Fig.\ref{fig:compare} shows the ratio of the energy gain possible between the super- and linear Fermi models for a plasmoid of a given size, assuming $\langle u_E \rangle_{pk}$ is constant for simplicity. Within each model, the total gain in energy $\Delta \varepsilon \doteq \varepsilon_\parallel/ \varepsilon_{||,0}$ is calculated for a particle which is allowed to drift inward until the island orbit width is $w=100 \rho_L$, where guiding center assumptions may weaken. In the linear Fermi calculation, $w$ is fixed to $w_p$, while for Super-Fermi Eq.\eqref{eq:evw} is used. The ratio of the total gain between the models $\Gamma \doteq \Delta \varepsilon_{SF}/\Delta \varepsilon_F$ is then shown as a function of the plasmoid size, here equivalent to the initial orbit width $w_0$. Consider an active region of the solar corona where $\rho_{L,e}\sim 0.1-1$ cm, alongside the relevant length scales of a solar flare~\cite{brooks21,chen2020}. The length of the current sheet itself is $\sim 10^9 \rho_{L,e}$, meaning the limiting ``monster" plasmoid size is still $w_p \approx 10^{7-8} \rho_{L,e}$~\cite{Loureiro2012,huang12}. Even for some of the much smaller more populous plasmoids, the Larmor scale separation present may allow $\Gamma > 10$, and accordingly a notable increase in energy gain over the linear Fermi model. In fact, given the asymptotic scalings $\Gamma \sim (w_p/\rho_L)^{2\chi}$ and $\Gamma_{UR} \sim (w_p/\rho_L)^{\chi}$, numerous more reconnection conditions such as active galactic nuclei disks and magnetars may also support similar increases in the maximum possible energy gain~\cite{ji11}. It should be stressed, however, that these are \textit{maximum} possible energy enhancements which we consider here. Many of the aforementioned examples with large Larmor scale separations are expected to possess a guide field, which is known to suppress Fermi acceleration~\cite{arnold2021,dahlin16}. Therefore, realistic gains in energy will likely not be as high. Additionally, these plasmoids would in practice have spatially dependent $u_E$, therefore the effective power law index of their acceleration could either be increased (likely for smaller populous plasmoids) or decreased (likely for the few ``monster'' plasmoids).
%
%
%

Although we only simulate test particles in anti-parallel reconnection here, we can at least make some predictions about the manner in which guide fields affect super-Fermi acceleration. In this model, guide fields are likely to play a role through the extension of various lengths out of the reconnection plane. The increase in radius of curvature and thus weakening of the curvature drift is countered exactly by an increase in path length in the acceleration region and hence energization time during a Fermi kick, leading to no change in $\Delta \varepsilon_\parallel$. On the other hand, the path length between Fermi kicks is extended to $w \rightarrow w\sqrt{1+(B_g/B_r)^2}$, modifying the denominator of Eq.~\eqref{eq:fermi1} accordingly. If $B_g/B_r$ were roughly constant or varied slowly, the super-Fermi power law becomes $p=1+\sqrt{1+(B_g/B_r)^2}/2\chi$. This, alongside the longer transit time between Fermi kicks, would cause steepening of power laws \textit{and} weakening of acceleration, mirroring expectations that Fermi acceleration is suppressed as described by Arnold et al~\cite{arnold2021} and Dahlin et al~\cite{dahlin16}. Perhaps in most cases however, it may be required to consider $B_g(w)/B_r(w)$ and integrate Eq.\eqref{eq:evw} exactly. The modification discussed above also does not consider the inherent change in plasmoid structure which may result from a guide field, such as a change to their pressure balance~\cite{sironi2016,majeski_2021}. Any changes to the plasmoid structure will likely carry over to the spatial/temporal dependence of $u_E$, and hence the effective observed power law of acceleration.

While figure \ref{fig:scatter} exhibits the constancy of $\chi$ in these nonlinear plasmoids, it is important to mention that there are circumstances where the constant-$\chi$ assumption does not appear to hold, like early on during the linear phase of plasmoid growth or during mergers. In these situations, $\chi$ in the left (differential) Eq.\eqref{eq:evw} will need to be considered more generally as $\chi(w)$, and the equation integrated accordingly to yield a different solution on the right. For linear plasmoids, the average curvature rises rapidly as field lines move away from the x-point. This results in a knee-like feature in $\dot{\varepsilon}_\parallel$ with small but steeply rising initial acceleration that rapidly levels off. This also occurs near x-points in nonlinear plasmoids, but only represents a transient compared to the power law phase.

Of paramount interest in any study of the acceleration of plasma particles is the energetic distribution such an acceleration mechanism would produce. Without any loss mechanisms and given a sufficiently low energy source, a constant acceleration rate of $\dot{\varepsilon} \sim \varepsilon^p$ yields an energetic particle distribution of $f(\varepsilon) \sim \varepsilon^{-p}$. MHD plasmoids are often considered to be the end of the road for energetic particles, suggesting that once trapped the particles have no means for exit. However in a dynamic current sheet such trapping is unlikely to last for the lifetime of a plasmoid including advection from the current sheet. Most plasmoids in a high Lundquist number current sheet will encounter multiple others with which they merge. A particle at the center of one plasmoid will, upon merger with a higher flux plasmoid, no longer be in the center and therefore experience continued acceleration~\cite{sironi2016}. Furthermore, a realistic 3D flux rope embedded in a current sheet is highly dynamic, with axial instabilities providing a prospect for inter-plasmoid transport of energetic particles~\cite{lgsk_19}. Lastly and most easily accounted for in this model is the fact that plasmoids are finite in the out of plane direction, either due to instability or a finite current sheet. Every $\Delta \varepsilon_\parallel$ in our model is accompanied by an axial step in the $z$ direction, which becomes larger as particles gain energy~\cite{dahlin_2020}. In fact, for reconnection rates $E$ much less than one, a particle may experience a larger relative change in its axial position than in its energy, meaning that axial transport is competitive with both trapping and energization. The loss of particles out of the ends of plasmoids and the rate of their re-injection would then serve as a cutoff in the particle energy for a single plasmoid, dependent on the reconnection rate.

\section{Conclusions}

We have proposed that an enhanced Fermi acceleration process exists in 2D multiscale MHD reconnection. Results from analytical theory and test particle simulations suggest that a correction arises from changing magnetic island orbit widths for particles. This yields an acceleration rate power law relationship $\dot{\varepsilon}_\parallel \sim \varepsilon_\parallel^{1.67}$ on average with the precise index varying somewhat due to island geometry, but generally remaining $\sim 1.5$ or larger. We additionally discussed the effects of the temporally- and spatially-varying $\vec{E}\times\vec{B}$ drift on the effective power law index, revealing a trend from high ($\gtrsim 2$) to low ($\lesssim 1$) power law indices as plasmoids get larger. In particular, this result places importance on the distribution of plasmoids in size and flux when investigating global particle energization in a multiscale current sheet~\cite{huang_bhattacharjee_2012,Loureiro2012,sironi2016,zhou2020}. 

Further generalization of this model would be most immediate with the development of a detailed understanding of the field and structure of plasmoid interiors as they grow~\cite{zenitani_miyoshi_2020}. Evolution of $u_E$ introduces a time dependence which would lead to a separable O.D.E. for the energy as a function of time, and hence a way to refine the power law of $\dot{\varepsilon}_\parallel(\varepsilon_\parallel)$. With knowledge of the spatial structure of these fields, Eq.\eqref{eq:evw} can once again be leveraged in directly modifying the power law. This would connect particle distributions to plasmoid distributions, possibly creating a route toward an analytical description of multiscale reconnection energetic particle spectra~\cite{majeski_2021,huang_bhattacharjee_2012}. Additionally, we assume particles are pre-energized by some injection mechanism which is likely beyond the scope of guiding center simulations.~\cite{ball19,Killian2020,French22,sironi22} Knowledge of the appropriate injection mechanisms for the current sheet we study would fix ideas about the efficiency of the combined processes of injection and Fermi acceleration, and produce a more complete model of particle acceleration in 2D multiscale MHD reconnection. Lastly, the equations used to evolve particles in this study only include terms up to first order in normalized Larmor radius ($\rho_L/L$).~\cite{northrop_61} Energization and cross-field transport resulting from higher-order finite Larmor radius effects can be captured by gyrokinetic models when Larmor scale separation is weak for the thermal plasma,~\cite{2006Howes} or particles have large initial $v_\perp$ (perhaps resulting from re-acceleration after escaping from another plasmoid and scattering off of an x-point~\cite{drake06}). Such effects will not be captured by the simplified guiding center system (equation \ref{eq:gce}).

Certain reconnection conditions will require non-trivial adjustments to this model in order to appropriately be described by it. In kinetic plasmas with smaller scale separation between Larmor radii and plasmoid widths, the original model of Drake et al may be more suited as particle motion is not so restricted to magnetic field lines, often due to $\mu$ being poorly conserved. The lack of $\mu$ conservation leads to \textit{both} $\varepsilon_\parallel$ and $\varepsilon_\perp$ increasing during energization events~\cite{guo_14}. Additionally, if a Hall-effect field is present then the curvature drift will have a projection in the plane of reconnection. Given that the curvature drift is charge dependent, this means that it will be directed out of the plasmoid for \textit{either} positive \textit{or} negatively charged particles~\cite{drake06, Drake2008, malyshkin08}, but not neither. This could impede the inward motion of the gyro-center due to the $\vec{E}\times\vec{B}$ drift, or even entirely reverse it as seen in Fig.2(b) of Drake et al (2006)~\cite{drake06}, yielding an opposite-sign correction to the Fermi rate power law. Whether this results in eventual escape from the island is uncertain however, and likely depends on details of the plasmoid structure and chaotic particle orbits which can occur near the x-points.

\vspace{1cm}
The data that support the findings of this study are
available from the corresponding author upon reasonable
request.
\begin{acknowledgements}
The authors thank Yi-Min Huang, Muni Zhou, and Nuno Loureiro for providing code and data for the MHD simulations, as well as Jim Drake, Fan Guo, and Matthew Kunz for helpful comments. Additional thanks are owed to the referees for their insightful suggestions which helped refine this paper. This work was supported by NASA grant no. 80HQTR21T0060 and DoE  Contract No. DE-AC02-09CH11466.
\end{acknowledgements}

\bibliography{fermi,rec}

\end{document}